\makeatletter
\def\ps@firstpage{\ps@plain
  \def\@oddfoot{}
  \def\@oddhead{\article@logo\hss}}
\def\article@logo{%
  \vbox to\headheight{%
    \@parboxrestore \@logofont
    \noindent
Tensor, N.~S.
    \newline Vol.~71 (2009), no.~2, p.~109--113.
    \par\vss
  }}
\def\@logofont{\fontsize{8}{9.6\p@}\selectfont}
\makeatother
\documentclass[twoside,leqno]{article}
\usepackage{amsfonts}
\usepackage[pdfstartview=FitH,pdfnewwindow=true,pdftitle={Physical significance of
(k²+A)|u| metric},pdfdisplaydoctitle=true,colorlinks,pagebackref,bookmarksopen,breaklinks]{hyperref}
\newcounter{tsection}
\newcommand{\tsection}[1]{\stepcounter{tsection}{\endgraf\medskip\normalfont\bfseries\hspace{.7ex}\S\hspace{.7ex}\thetsection.\hspace{.8em}#1\ }}
\newcommand{\tintroduction}[1]{{\endgraf\normalfont\bfseries\hspace{.7ex}#1\ }}
\newcommand{\msp}{{\kern.03em}}
\newcommand{\ga}[3]{{\Gamma^{#1}{}_{#2#3}}}
\newcommand{\bu}{{u}}
\newcommand{\bw}{{{u'}}}
\newcommand{\bp}{{\pi}}
\newcommand{\bq}{{\pi^{(1)}}}
\newcommand{\bpr}{{{'}}}
\newcommand{\pr}[2]{{{#1\!\cdot\!#2}}}
\newcommand{\nbu}{{\left\|{u}\right\|^{\mathstrut}}}
\newcommand{\nw}[2]{{\left\|{#1}\wedge{#2}\right\|}}
\newcommand{\N}[1]{{\left\|{u}\right\|^{\mathstrut\scriptscriptstyle#1}}}
\newcommand{\sss}[1]{^{\mathstrut\scriptscriptstyle#1}}
\let\dfrac\frac
\newtheorem{PRP}{Proposition}
\let\Tarraycolsep\arraycolsep\newlength{\TArraycolsep}\TArraycolsep
\newbox\kmetric\setbox\kmetric=\hbox{ $(k^{2}+A)\nbu$ }
\begin{document}
\title{\renewcommand{\thefootnote}{\fnsymbol{footnote}~}\footnotetext{Received January 5, 2009}TOWARDS THE PHYSICAL SIGNIFICANCE OF THE $(k^{\sss2}+A)\nbu$ METRIC.\footnotemark
     \footnotetext{The work was presented at the 10\textsuperscript{th} International Conference of Tensor Society held at Constan\c ta, Romania, Sept.~3--7, 2008.}
\kern2pt\footnotemark\footnotetext{Research supported by the grant GA\v CR 201/09/0981 of the Czech Science Foundation.}
}
\renewcommand{\thefootnote}{\arabic{footnote})}
\author{By \href{http://iapmm.lviv.ua/12/eng/files/st_files/matsyuk.htm}{Roman \sc Matsyuk}
}
\date{}
\maketitle
\thispagestyle{firstpage}
\begin{abstract}
We offer an example of the second order Kawaguchi metric function the extremal flow of which generalizes the flat
space-time model of the semi-classical spinning particle to the framework of the pseudo-Riemannian space-time.
The general shape of the variational Euler--Poisson equation of the fourth order in the (pseudo-)Riemannian space
is being developed too.
\end{abstract}

\tintroduction{Introduction.} In 1946 Fritz Bopp in an attempt to describe the relativistic motion of the charged
particle influenced by self-radiation in flat space-time considered a Lagrange
function~\cite{matsyuk:Bopp1946}\footnote{Numbers in brackets refer to the references at the end of the paper.},
which, in the absence of the external electromagnetic field, may be expressed in terms of the particle's world
line Frenet curvature as follows:
\begin{equation}\label{maciuk:L}
  L^k=(k^{\sss2}+A)\nbu\,,
\end{equation}
where $u$ denotes the derivative $\dot x$ of the configuration space variable~$x$ with respect to the evolution
parameter~$\xi$ along the particle's world line~$x^n(\xi)$. Later different modifications of Bopp Lagrangian were
introduced, among them a more general expression was investigated by Lovelock in
1963~\cite{matsyuk:Lovelock}.\footnote{We even do not attempt to present here an exhaustive bibliography on the subject.} Then, in 1972, Riewe, still staying in the framework of flat space-time, proposed an equation of the
fourth order with the purpose to give a description of the semi-classical ``Zitterbewegung'' of test particle with
an internal degree of freedom:
\begin{equation}\label{maciuk:Riewe}
  \frac{d^{\sss4}x^n}{ds^{\sss4}}+\omega^{\sss2}\frac{d^{\sss2}x^n}{ds^{\sss2}}=0\,,
\end{equation}
where the derivatives are calculated with respect to the natural parameter. Recently in papers \cite{matsyuk:NAS}
and \cite{matsyuk:Zitt} I showed that the Riewe equation follows from the Bopp Lagrangian under the a posteriori
imposed constraint~$k^{\sss2}={1\over3}\,A+{2\over3}\,\omega^{\sss2}$. The goal of the present communication is to
obtain a generalization of the equation~(\ref{maciuk:Riewe}) from the variational principle with the fundamental
function~(\ref{maciuk:L}) in the (pseudo-)Riemannian case. The space, endowed with the metric
function~(\ref{maciuk:L}), may be considered as an example of a Kawaguchi space, because this function~$L^k$
satisfies Zermelo conditions.

\tsection{The covariant momenta.} Let us introduce the following change of local coordinates in the second-order
velocities space:
\[\{x^n,u^n,\dot u^n\}\mapsto\{x^n,u^n,u'\msp^n\}\,,\]
where the prime stands for the covariant derivative. Let us also denote the local expression of the Lagrange
function in terms of the new coordinates by~$\tilde L$. The following formul{\ae} produce then the receipt of the
recalculation of partial derivatives:
\begin{equation}\label{from u dot to u'}
  \frac{{\partial
L}}{\partial u^n}=\frac{\partial \tilde L}{\partial u^n}+2\,\frac{\partial \tilde L}{\partial u'\msp^q}\ga q m
nu^m,\qquad \frac{{\partial L}}{\partial x^n}=\frac{{\partial \tilde L}}{\partial x^n}+\frac{\partial \tilde
L}{\partial u'\msp^q}\,\frac{\partial\ga q m l}{\partial x^n}\,u^lu^m\,.
\end{equation}

For further use we recall the familiar conventions from the Riemannian geometry
\begin{eqnarray}\label{matsyuk:AppCov}
        &\displaystyle a'\,^n=\frac{da^n}{d\xi}+\ga n l m a^m u^l , \qquad a'{}_n=\frac {da_n}{d\xi}-\ga m l n a_m u^l \,,&\\
\label{matsyuk:AppDxG}
        &\displaystyle \dfrac{\partial g_{mn}}{\partial x^k}=g_{ml}\ga l k n +g_{nl}\ga l k m \,,&\\[1\jot]
\label{matsyuk:AppR}
        &\displaystyle R_{kmn}{}^l=\dfrac{\partial \ga l k n}{\partial x^m}
        -\dfrac{\partial \ga l m n}{\partial x^k}+\ga l m q \ga q k n
        -\ga l k q \ga q m n \,.&
\end{eqnarray}

Let us introduce the covariant momenta
\begin{equation}
 \bq =\dfrac{\partial \tilde L}{\partial \bw}\,,\qquad
\label{matsyuk:pi1}
 \bp =\dfrac{{\partial \tilde L}}{\partial \bu}-\bq\bpr \,.
\end{equation}
\begin{PRP}\label{maciuk:mainPRP}
Let some Lagrange function~$L$ depend on all the variables exclusively through the differential invariants
$\gamma=\pr uu$, $\beta=\pr u{u'}$, and $\alpha=\pr{u'}{u'}$ only. In this case the Euler--Poisson expression is:
\begin{equation}
  \label{matsyuk:E II ultimate}
\fbox{$\displaystyle{\cal E}_n=-\pi'{}_n-\pi^{(1)}{}_lR_{nkm}{}^lu^mu^k$}
\end{equation}\end{PRP}
{\sl The proof is given in steps:}
\newcounter{PRPcounter}\newbox\STEP\setbox\STEP\hbox{\sl \hspace{2em} Step}
\begin{list}{\sl Step~\arabic{PRPcounter}\/.\hfill}{\usecounter{PRPcounter}\leftmargin0pt
\setlength{\itemindent}{\wd\STEP}
}
\item
In second order Ostrohrads\kern-.15em'kyj mechanics the Euler--Poisson expression~${\cal E}$, that constitutes the
system of variational Euler--Poisson equations~$\{{\cal E}_n=0\}$ is known to be conveniently put down in terms of
the momenta

\begin{equation}
  \label{matsyuk:p} p^{(1)}_n=\dfrac{\partial L}{\partial\dot u^n}\,,\qquad p_n=\dfrac{\partial L}{\partial
u^n}-\frac{dp^{(1)}_n}{d\xi}\,,
\end{equation}
as follows
\begin{equation}\label{matsyuk:E-P}
{\cal E}_n=\dfrac{\partial L}{\partial x^n}-\dfrac{d p_n}{d\xi}=0\,.
\end{equation}
\item
The covariant momentum~$\bp$, profiting from the first of the formul{\ae}~(\ref{from u dot to u'}) together with
the covariant derivative pattern~(\ref{matsyuk:AppCov}), is presented as:
\begin{equation}
{\bp}_n =\dfrac{\partial L}{\partial u^n}-2\,\ga q m n u^m \pi^{(1)}{}_q
 -{\pi^{(1)}}\msp'{}_n\,.
 \label{matsyuk:pi}
\end{equation}
The covariant derivative of the momentum~$\bq$, again profiting from the pattern~(\ref{matsyuk:AppCov}), writes
down as
\begin{equation} \label{matsyuk:pi1'}
  {\pi^{(1)}}\msp'{}_n=\frac d{d\xi}\pi^{(1)}{}_n-\ga m l n \pi^{(1)}{}_m u^l \,.
\end{equation}
\item
in terms of the covariant quantities above, the non-covariant quantity~$p_n$ from the expression~(\ref{matsyuk:p})
is given by the following calculation:
\begin{eqnarray}
  p_n  &=  & \pi_n+2\,\ga q m n u^m\pi^{(1)}{}_q+\pi^{(1)}\msp'{}_n-\phantom{\frac d{d\xi}\pi^{(1)}{}_n}\quad\mbox{\rm (by virtue of (\ref{matsyuk:pi}))} \nonumber \\
   &  & \phantom{\pi_n+2\,\ga q m n u^m\pi^{(1)}{}_q+\pi^{(1)}\msp'{}_n}-\frac d{d\xi}\pi^{(1)}{}_n \quad\mbox{\rm \hspace{.5em}(by virtue of (\ref{matsyuk:pi1}))}
   \nonumber\\[2\jot]
   & = &\pi_n+\ga q m n u^m\pi^{(1)}_q \qquad\qquad\mbox{\rm (by virtue of (\ref{matsyuk:pi1'})).}\label{matsyuk:p_n}
\end{eqnarray}
Differentiating~(\ref{matsyuk:p_n}) and applying the pattern~(\ref{matsyuk:AppCov}) in order to express the
ordinary derivatives of the variables $\pi$ and $u$ in terms, respectively, of the covariant derivatives $\bp\bpr$
and $\bu\bpr$, and implementing the guise~(\ref{matsyuk:pi1'}), produces:
\begin{eqnarray*}
 \frac d{d\xi} p_n&=&\big(\pi'{}_n+\ga l m n \pi_l u^m\big)+\dfrac{\partial \ga l
m n}{\partial x^k}
 u^ku^m\pi^{(1)}{}_l \\
&& {}+\big(\ga l m n {u'}\msp^m-\ga l m n
 \ga m q k u^qu^k\big)\pi^{(1)}{}_l
 +\ga l m n u^m\big({\pi^{(1)}}\msp'{}_l
 +\ga q k l \pi^{(1)}{}_q u^k\big)  \\[2\jot]
&=&\pi'{}_n+\big(\pi^{(1)}\msp'{}_l+\pi_l\big)\ga l m n u^m+\pi^{(1)}{}_l\ga l m n u'\msp^m \\[2\jot]
&&{}+\pi^{(1)}{}_q u^m u^k\left(\ga l m n \ga q l k+\dfrac{\partial \ga q m n}{\partial x^k}
        -\ga q l n \ga l m k\right).
\end{eqnarray*}
\item
Now the Euler--Poisson expression~(\ref{matsyuk:E-P}) takes on the shape
\[
{\cal E}_n=\dfrac{\partial \tilde L}{\partial x^n}
        -\big(\pi^{(1)}\msp'{}_l+\pi_l\big)\ga l m n u^m
        -\pi^{(1)}{}_l\ga l m n u'\msp^m
        -\pi'{}_n-\pi^{(1)}{}_l u^mu^kR_{nkm}{}^l \,.
\]
Let us show, that the first four addends in this expression produce zero,--- under the assumptions of the
proposition we  are now proving. For the sake of constructing the expression
\begin{equation}\label{maciuk:dL/dx}
\frac{\partial \tilde L}{\partial x^n}=\frac{\partial \tilde L}{\partial\gamma}\frac{\partial \gamma}{\partial
x^n}+\frac{\partial \tilde L}{\partial\beta}\frac{\partial\beta}{\partial x^n}+\frac{\partial \tilde
L}{\partial\alpha}\frac{\partial\alpha}{\partial x^n}\,,
\end{equation}

using formula~(\ref{matsyuk:AppDxG}), we calculate:\[\frac{\partial \gamma}{\partial x^n}=2\ga
lmnu^mu_l\,,\quad\frac{\partial \beta}{\partial x^n}=\ga lmnu^mu'{}_l+\ga lmnu'\msp^mu_l\,,\quad\frac{\partial
\alpha}{\partial x^n}=2\ga lmnu'\msp^mu'{}_l\,.\] On the other hand, applying the definitions~(\ref{matsyuk:pi1}),
we get:
\begin{equation}\label{maciuk:MOMENTA}
\pi^{(1)}_n=\frac{\partial \tilde L}{\partial \beta}u_n+2\frac{\partial \tilde L}{\partial
\alpha}u'{}_n\,,\quad\pi^{(1)}\msp'{}_n+\pi_n=2\frac{\partial \tilde L}{\partial \gamma}u_n+\frac{\partial \tilde
L}{\partial \beta}u'{}_n\,.
\end{equation}
Extracting these two expressions from~(\ref{maciuk:dL/dx}) produces zero.\hfill $\square$
\end{list}

\tsection{The generalized variational equation of a structured particle in Riemannian space.} Now it is
straightforward to obtain the equation of the extremal world line for the model~(\ref{maciuk:L}). Recalling the
expression of the first Frenet curvature,
\[k=\frac{\nw u{u'}}{\N3}\,,\] one sees that in terms of the invariants $\gamma$, $\beta$, and $\alpha$, the Lagrange
function~(\ref{maciuk:L}) takes the shape
\[L^{k}=\frac{\alpha\gamma-\beta^{2}}{\gamma^{5/2}}+A\gamma^{1/2}\,,\]
from where by means of the formul{\ae}~(\ref{maciuk:MOMENTA}) together with the differential prolongation of the
first of them,
\[\pi^{(1)}{}'{}_n=\left(\frac{d}{d\xi}\frac{\partial \tilde L}{\partial \beta}\right)u_n
+\frac{\partial \tilde L}{\partial \beta}u'{}_n
    +2\left(\frac{d}{d\xi}\frac{\partial \tilde L}{\partial \alpha}\right)u'{}_n
+2\frac{\partial \tilde L}{\partial \alpha}u''{}_n\,, \] one immediately obtains:
\begin{eqnarray}\label{maciuk:P1ultimate}
     \pi^{(1)}&=&\frac2{\N3}\,u'-\,\frac{2\,\pr u{u'}}{\N5}\,u\,,  \\[1\jot]
    \pi&=&\left(\frac{2\,\pr u{u''}}{\N5}-\frac{\pr {u'}{u'}}{\N5}-\frac{5\,(\pr u{u'})^2}{\N7}+\frac{A}{\nbu}\right)u  + \frac{6\,\pr u{u'}}{\N5}\,u'-\frac{2}{\N3}\,u''. \label{maciuk:Pultimate}
\end{eqnarray}
\tsection{Relation to physics.}
\vspace*{-2\jot}
\paragraph{The Riewe equation.}
The Euler--Poisson equation~(\ref{matsyuk:E II ultimate}) for~$L^k$ inherits the property of parametric
ambivalence from the same property of the corresponding variational problem with the fundamental
function~(\ref{maciuk:L}) due to the fulfillment of the Zermelo conditions. Thus it is possible to pass to the
natural parametrization by the arc length~$s$, $u_nu^m=1$, in the expression~(\ref{maciuk:Pultimate}), while substituting it in~(\ref{matsyuk:E II
ultimate}). Then one gets
\[\frac D{ds}\left[\left(-3\,\pr{u'_s}{u'_s}+A\right)u_n-2\,(u''_s)_n\right]=-\pi^{(1)}{}_lR_{nkm}{}^lu^mu^k\,.\]
The Riewe equation~(\ref{maciuk:Riewe}) follows from this expression in flat space-time on the surface $k=\rm
const$.
\paragraph{The Dixon equations.} General relativistic top with inner angular momentum~$S^{nm}$ in
pseudo-Riemannian space-time is in common knowledge described by means of the system of first-order
equations~\cite{matsyuk:Dixon}\footnote{The definition of the curvature tensor, adopted in the present
communication, differs in sign from the one used in the Dixon's paper}
\begin{equation}\label{maciuk:Dixon}
  \left\{\begin{array}{rcl}
    P'{}_n & = & -\displaystyle{1\over2}\,R_{nm}{}^{kl}u^mS_{kl}\,, \\[2.5\jot]
  S'{}_{nm} & = & \phantom{-}P_nu_m-P_mu_n\,.
  \end{array}\right.
\end{equation}
By the skew-symmetric property of the Riemannian curvature tensor it easily follows that the first of the above
equations is regained by putting $P=\bp$ and $S=u\wedge\bq$ in~(\ref{matsyuk:E II ultimate}).
\begin{PRP}Under the assumptions of the Proposition~\ref{maciuk:mainPRP} the governing system of
equations~(\ref{maciuk:Dixon}) does not depend on any particular appearance of the fundamental function~$L$.
\end{PRP}
This follows from formul{\ae}~(\ref{maciuk:MOMENTA}) along with the similar formula for~$\bp$.\hfill $\square$

\bigskip
\newbox\instit\setbox\instit=\hbox{in Mechanics and Mathematics}
\newdimen\minipg\minipg\wd\instit
\rightline{\begin{minipage}{\minipg}\begin{center}
Institute for Applied Problems\\in Mechanics and Mathematics\\15~Dudayev~St.\\290005 L\kern-1pt'viv, Ukraine
\\E-mail: matsyuk@lms.lviv.ua, romko.b.m@gmail.com
\end{center}\end{minipage}}

\end{document}